\documentstyle[epsfig]{aipproc}

\def\be#1\ee{\begin{equation}#1\end{equation}} 
\def\bea#1\eea{\begin{eqnarray}#1\end{eqnarray}}

\def                            
  \Complexes
   {{\rm C}\llap{\vrule height6.3pt width1pt depth-.4pt\phantom t}}

\def\=>{\Rightarrow}

\def\D{{\rm{D}}}

\def\sqr#1#2{{\vcenter{\vbox{\hrule height.#2pt
         \hbox{\vrule width.#2pt height#1pt \kern#1pt
            \vrule width.#2pt}
         \hrule height.#2pt}}}}

\def\BulletItem #1 {\item{$\bullet$}{#1}}  
\def\Integers{{\bf Z}}
 
\def\=>{\Rightarrow}
\def\==>{\Longrightarrow}
 
 \def\dal{\displaystyle{{\hbox to 0pt{$\sqcup$\hss}}\sqcap}}
 
\def\lto{\mathop
        {\hbox{${\lower3.8pt\hbox{$<$}}\atop{\raise0.2pt\hbox{$\sim$}}$}}}
\def\gto{\mathop
        {\hbox{${\lower3.8pt\hbox{$>$}}\atop{\raise0.2pt\hbox{$\sim$}}$}}}
\def\bar{\overline}		
\def\hat{\widehat}		

\def\Reals{{\rm I\!\rm R}}	

\def				
  \Complexes
   {{\rm C}\llap{\vrule height6.3pt width1pt depth-.4pt\phantom t}}

\def\tensor{{\otimes}}		

\def\interior #1 {  \buildrel\circ\over  #1}     

\def\semidirect{\mathbin
               {\hbox
               {\hskip 3pt \vrule height 5.2pt depth -.2pt width .55pt 
                \hskip-1.5pt$\times$}}}

\def\S{{\cal S}}                
\def\R3{\Reals^3}
\def\RP3{\Reals P^3}

\newcommand{\blist}{\begin{list}{}{\setlength{\leftmargin}{4mm}
\setlength{\parindent}{0mm}\setlength{\parsep}{1mm}
\setlength{\topsep}{2mm}}}
\newcommand{\elist}{\end{list}}

\begin{document}
\title{Spin and Statistics in Quantum Gravity} 

\author{H.F. Dowker$^*$ and R.D. Sorkin$^{\dagger}$}
\address{$^*$Department of Physics,
Queen Mary and Westfield College,\\
 London E1 4NS, UK.\\
$^{\dagger}$Department of Physics,
Syracuse University,\\
Syracuse, NY 13244-1130, USA.
        } 

\maketitle

\begin{abstract}
We present a review of the spin and statistics of topological geons, 
particles in $3+1$ quantum gravity. They can have half-odd-integral
spin and fermionic statistics and  since the 
underlying gravitational field is tensorial and bosonic, this
is an example of ``emergent'' non-trivial spin and statistics as
displayed by familiar non-gravitating objects such as 
skyrmions. We give the topological background  
and show that  
in a ``canonical'' quantization of 
gravity there is no spin-statistics
correlation for topological geons. Allowing the
topology of space to  change, for example in a sum-over-histories
approach, raises the possibility that a spin-statistics correlation 
can be recovered
for geons. We review a 
conjectured  set of rules powerful enough to give such a 
spin-statistics correlation  
for all topological geons. These would appear to
rule out the possibility 
of parastatistics and may rule out spinorial and 
fermionic geons altogether.

\end{abstract}


\section{Introduction}

In quantum mechanics and in quantum field theory in flat spacetime
we have many examples of 
``emergent'' fermionic statistics and spinorial
({\it i.e.}, half-odd-integral) spin for objects built from entities which are
fundamentally tensorial ({\it i.e.}, integral spin) and bosonic. 
Can such a phenomenon occur in quantum gravity in which the  
dynamical variable, the spacetime
metric, is tensorial and bosonic? The answer is yes, 
and in this review we will look at the best studied case, that of 
so-called ``topological geons'', particles made of
non-trivial spatial topology, in $3+1$ dimensions.

The original work on spin-half states in quantum 
gravity  was done by Friedman and 
Sorkin \cite{Friedman:1980st,Friedman:1982du} and on 
fermionic states by Sorkin  \cite{Sorkin:1985bg}.
In section II we recall the basic features of topological geons,
particles which exist by virtue of the non-trivial 
topology of space, and 
show how they acquire their spin and statistics by a choice of 
unitary irreducible representation (UIR) of the so-called
mapping class group (MCG). 
We will see that there is no spin-statistics correlation for quantum geons
\cite{Sorkin:1985bg,%
Sorkin:1988mw,Aneziris:1989xz,Aneziris:1989cr} 
and indeed there are always spin-statistics violating 
sectors for any species of geon \cite{Sorkin:1996yt,Sorkin:1996cv}. 
Moreover, the presence of a particular type of symmetry between quantum geons,
namely the ``slides'' which correspond to diffeomorphisms in which 
one geon slides through another, produces an extraordinary variety of 
quantum sectors \cite{Sorkin:1996yt,Sorkin:1996cv}. 
 
The lack of a spin-statistics correlation for geons is perhaps not surprising.
In the proofs of
all existing spin-statistics theorems for extended
objects like geons,  the 
possibility of particle-anti-particle creation (and annihilation) is 
crucial.  Indeed particle-anti-particle pair
production and annihilation 
has been suggested (see \cite{Balachandran:1990wr,Balachandran:1993mf})
as a unifying principle that might bring 
together the  
``topological'' 
spin-statistics theorems, of which the Finkelstein-Rubenstein
version \cite{Finkelstein:1968hy} is the original, and the 
relativistic quantum field theory theorems (the conditions imposed 
--- Lorentz invariance {\it {etc.}} --- being 
supposedly just those that guarantee that antiparticles exist
with the required possibilities for pair creation and annihilation). 

Now, the process of geon-anti-geon pair production is a topology
changing one and cannot be described within a formalism
like canonical quantum gravity which assumes, a
priori, that the spatial three-manifold is fixed. It has therefore been
conjectured  that in a formulation of quantum gravity which can
accommodate topology change, the usual spin-statistics connection would be
recovered for geons \cite{Sorkin:1988mw}. 

The sum-over-histories (SOH) for quantum gravity is such a formulation
and there is indeed evidence that there's a spin-statistics theorem
for geons ``trying to get out'' in a SOH approach \cite{Dowker:1996ei}.
To prove a general spin-statistics theorem for all geons, 
extra assumptions are needed and in section III we review a set of rules
which would achieve this \cite{Sorkin:1988mw}. The consequences
of these rules are stronger than originally envisaged and there's
a possibility that they might rule out spinorial and fermionic
geons altogether.

Section IV is a brief mention of some motivation for pursuing a 
geon spin-statistics theorem. 
 
We will
restrict ourselves to orientable three-manifolds throughout
and 
will further assume
that no handles, $S^2 \times S^1$, occur in the ``prime decomposition'' of
the three-manifold (see section II).

\section{How the geon got its spin and statistics}

Topological geons are particles made from non-trivial spatial
topology.  We are interested in the situation of an isolated system of
these 
particles and thus we will be dealing with a three-dimensional manifold $M$
which admits asymptotically flat metrics.  Physically, $M$ is three-space
at a ``moment of time'', or, the ``future boundary of
truncated spacetime'' \cite{Sorkin:1994cd}.

\subsection{The spatial three-manifold}

There is a ``Three-Manifold Decomposition Theorem'' that identifies
candidates for elementary geons, but in order to state this theorem we must
first introduce the concepts of ``connected sum''
(denoted $\#$) and ``prime manifold.''
To take the connected sum of two oriented three-manifolds $M_1$ and $M_2$,
remove an open ball from each and identify the resulting two-sphere
boundaries with an orientation-reversing diffeomorphism (henceforth,
``diffeo'').  Taking the connected sum of any three-manifold with $S^3$
gives a manifold diffeomorphic to the original one; taking it with $\Reals^3$
is topologically equivalent to deleting a point.  A prime three-manifold,
$P$, is a closed three-manifold that is not $S^3$ and such that whenever
$P=P_1\# P_2$, either $P_1$ or $P_2$ is $S^3$. Examples of primes are the
three-torus, $T^3$, and the so-called spherical spaces, $S^3/G$, where $G$
is some discrete subgroup of $SO(4)$ acting freely on $S^3$.

The $M$ we are considering is $M=\Reals^3 \# K$ where K is a closed
three-manifold.  The Decomposition Theorem states that any such $M$ can be
decomposed into the connected sum of finitely many prime manifolds and this
decomposition is unique:
\be
M=\Reals^3 \# P_1 \# P_2 \dots \# P_n.
\ee
We will assume that to each prime summand there corresponds an elementary
quantum geon; with ``correspond'' being used in a suitable sense since
there is a rather subtle relation between a particular piece of spatial
topology and a physical particle---which subtlety has to do both with
familiar ``identical particle exchange effects'' and unfamiliar effects due
to the existence of diffeos known as ``slides'' \cite{Sorkin:1985bg,%
Sorkin:1988mw}.


 For more details see \cite{Sorkin:1985bg,Sorkin:1988mw,Friedman:1988qs}

\subsection{Wave functions}

In canonical quantum gravity, for which the topology does not change, the
configuration space, $Q$, is the space of all three-geometries on $M$,
\be
Q= {{{\rm{Riem}}^\infty(M)}\over{{\rm{Diff}}^\infty(M)}},
\ee
where ${{\rm{Riem}}^\infty(M)}$ (${\rm{R}}^\infty$ for short) is the space
of asymptotically flat Riemannian metrics on $M$ and
${{\rm{Diff}}^\infty(M)}$ ($\D^\infty$ for short) is the group of
diffeomorphisms of $M$ that become trivial on approach to
infinity.
It can be shown that ${\D}^\infty$ acts freely on ${\rm{R}}^\infty$ and so
$Q$ is a manifold, ${\rm{R}}^\infty$ being a 
principal fibre bundle over $Q$ with
fibre $D^\infty$.  Thus, using the fact that ${\rm{R}}^\infty$ 
is convex and hence
contractible to a point so that all its homotopy groups are trivial, we
deduce that $\pi_k(D^\infty)\simeq\pi_{k+1}(Q)$.

Wave functions need not be single-valued on $Q$ if $Q$ contains
non-contractible loops.  Rather, the transformation of a wave function as
such loops are traversed gives a representation of $\pi_1(Q)$.  This is a
special case of the general situation where wave functions are sections of
a twisted vector bundle on $Q$. Physical observables are 
invariant under ${{\rm{Diff}}^\infty(M)}$
which means that the quantum state space 
reduces to a number of 
inequivalent and independent sectors each transforming under
a different unitary irreducible representation 
of the group $\pi_1(Q)$. From the above we 
know that $ \pi_1(Q) \simeq \pi_0(D^\infty)$, where $G \equiv\pi_0(D^\infty)$
is known as the mapping class group (MCG) or the group of
large diffeos of $M$ and is the analogue for gravity of the group 
of large gauge transformations in a gauge theory. 

\subsection{The mapping class group and UIR's}

Let us restrict attention to the manifold which is 
a connected sum of $\Reals^3$ and a number, $n$, of identical primes, 
$M = \Reals^3\# P\#\dots  P$.
$G$ is generated
%
%
by three types of large diffeomorphism, 
the {\it exchanges}, the {\it
internal diffeomorphisms} and the {\it slides}.  
Each of these generators can be viewed as the result of a certain {\it
process} (a ``development'' \cite{Sorkin:1985bg}), with the nature of the
process being suggested by the name of the category.  Thus, an exchange is
the result of a process in which
two 
identical primes continuously change their positions until they have 
swapped places.
Similarly a slide is the result of a process in which one prime travels
around a closed 
loop threading through one or more other primes, while an internal
diffeo is a diffeo whose support is restricted to a single
prime. 

One very important internal diffeo is the $2\pi$-rotation of a single prime. 
It is the result of a process in which the prime rotates around by 
$2\pi$. For particular primes, $P$, this diffeo is deformable to the identity 
and so is not a non-trivial element of $G$. Geons based on these primes
must have tensorial (integral) spin since a $2\pi$ rotation is equivalent 
to the identity and so cannot have a non-trivial effect on the quantum 
state. Of the known primes, only the handle (which 
we are explicitly excluding here) and the lens spaces 
are tensorial: the rest have non-trivial $2\pi$ rotation and are called
spinorial primes.  

The slides, internals and exchanges each form subgroups of $G$ and 
the exchanges generate a subgroup isomorphic to the 
permutation group, $\S_n$.
In \cite{Sorkin:1996yt,Sorkin:1996cv} the literature on the MCG 
was sublimated into the following result
\be
\label{semi.eq} 
   G = (slides) \semidirect (internals) \semidirect \S_n  ,
\ee
where the symbol $\semidirect$ denotes semidirect product (with the normal
subgroup on the left).  What this says is that every
element of $G$ is uniquely a product of three diffeomorphism-classes, one
from each subgroup, and that each subgroup is invariant under conjugation
by elements of the subgroups standing to its right in equation 
\ref{semi.eq}.
See 
\cite{Giulini:1994de,Giulini:1995ui,Giulini:1997hi} for 
more details on the MCG. 

The fact that $G$ is a semidirect product allows us 
to analyze its UIR's in
terms of representations of its factor groups and their subgroups.
Let
\be
      G = N \semidirect K
\ee
be a semidirect product with $N$ being the normal subgroup.  A finite
dimensional UIR of $G$ is then determined by the following data
{\obeylines{
 $\bullet$ \ $\Gamma$ \ = a UIR of $N$
 $\bullet$ \ $T$ \ = a PUIR of $K_0 \subseteq K$ ,}}

\noindent
where $K_0$ is the subgroup of $K$ that remains ``unbroken by $\Gamma$''
($K_0$ is ``the little group'') and ``PUIR'' stands for ``projective UIR'',
{\it {i.e.}}, representation up to a phase. (The Schur multiplier 
for $T$ is determined by $\Gamma$.)


In seeking the UIR's of this group there are two situations,
depending on whether the slide subgroup is represented trivially or not.
The results in the following two subsections are contained in 
\cite{Sorkin:1996yt,Sorkin:1996cv}.

\subsection{The sectors with trivial slides}
\label{trivslides.subsection}

In the simpler case of UIR's of $G$ which annihilate the slides, 
in effect a complete classification is possible. 
In this case, the mathematical
problem is reduced to finding the UIR's of the quotient group,
$G/(slides)$, which by equation \ref{semi.eq} is just the semidirect product
\be{\label{noslides.eq}}
   (internals) \semidirect \S_n .      
\ee
Let us find the finite dimensional UIR's of this group (sometimes called
the ``particle group'' 
\cite{Giulini:1997hi}). 
The normal group of internal diffeos is the direct
product of $n$ copies of the MCG group, $H$, for a single
prime $P$:
\be
   N = (internals) = H \times H \times \cdots \times H . 
\ee
The most general UIR of
this normal subgroup $N$ is itself a product, namely the tensor product,
\be\label{tprod.eq}
    \Gamma = \Gamma_1 \tensor \Gamma_1 
              \cdots 
             \Gamma_2\tensor\Gamma_2
              \cdots 
             \Gamma_r ,                        
\ee
of UIR's of $H$, 
where the first $n_1$ factors are $\Gamma_1$, the next $n_2$ factors
are $\Gamma_2$, {\it {etc.}}.
Physically a given $\Gamma_i$
specifies a certain ``internal structure'' for the corresponding geon, and
is therefore a ``species parameter'' or ``quantum number''.
Different $\Gamma_i$'s mean that the corresponding geons
are not identical particles but are of different physical species.
So here can we have a quantum breaking of indistinguishability. 

With respect to the choice (\ref{tprod.eq}),
the unbroken subgroup
$K_0\subseteq\S_n$ reduces to a product of permutation groups,
\be\label{little.eq}
    K_0 = \S_{n_1} \times \S_{n_2} \times \cdots \times \S_{n_r} .
\ee
The statistics is then given by a UIR $T$ of
$K_0$, that is to say by an independent UIR $T_1$, $T_2,\cdots$ for each
of the subgroups $\S_{n_1}$, $\S_{n_2}, \cdots$.  Each of these $T$'s in
turn, can be specified by a choice of a Young tableau, and determines
whether the corresponding geons will manifest Bose statistics, Fermi
statistics or some particular parastatistics.  Since there is no
restriction on the choice of $T$, there is no restriction on which
combinations of these possible statistics can occur.

To summarize, all possible
sectors with trivial slides are accounted for by specifying%
{ \obeylines{
$\bullet$\  { a {\it species} for each geon (i.e. a UIR of $H$) }
$\bullet$\  { a {\it statistics} for each resulting set of identical geons }
}}

\subsection {Some sectors with nontrivial slides}

When the slide subgroup is represented nontrivially, 
a full classification of the possible UIR's of $G$ does not exist.
It is clear, however, that there is an extraordinary variety 
of possible UIR's
and in order to give an idea  of this richness
we give an example of  a UIR in  
a special case which avoids most of the
complications which obstruct the full classification. 

The prime in this special case 
is $\RP3$, which can be visualized as a region of $M$
produced by excising a solid ball and then identifying antipodal pairs of
points on the resulting $S^2$ boundary.  The internal group is
trivial for this prime.  
For each pair of $\RP3$'s, one can slide one through the other, with the
square of this slide being trivial (since $\pi_1(\RP3)=\Integers_2$),
making a total of $n(n-1)$ independent order 2 generators.  The complete
group $(slides)$ is then generated by products of these elementary slides.

Since for $\RP3$, $(internals)$ is trivial, the MCG reduces to
\be
          G = (slides) \semidirect \S_n .
\ee
Hence, according to the general scheme outlined earlier, we get a UIR of
$G$ by choosing first a UIR, $\Gamma$, of $N=(slides)$, and then a 
suitable PUIR, $T$,
of the resulting unbroken
subgroup $K_0\subseteq{}\S_n$.  As before, we may interpret $K_0$ as
describing the surviving indistinguishability of the geons, and $T$ as
describing the statistics within each set of identical geons.  

Consider abelian UIRs of $N$ in which, in 
effect, each ordered pair of primes is assigned a $\pm 1$. 
We can represent the various such UIR's pictorially by drawing $n$ 
dots to represent the primes and an arrow to represent each ordered
pair that receives a minus sign (meaning the a slide of the first prime
through the second produces a phase-factor of $-1$).  Each distinct diagram
of this type then gives rise to a different class of UIR's of $(slides)$,
and therefore furnishes a different building block for constructing UIR's
of $G$.

If we make a particular choice of UIR of $(slides)$ by  
choosing a diagram in which the dots and arrows form a circle
and the arrows point anticlockwise, say,
this pattern leaves $\Integers_n\subseteq\S_n$ as the unbroken subgroup
$K_0$. We acquire distinct UIR's of $G$ corresponding to the
possible UIR's of $\Integers_n$. Here, 
geon identity is expressed not by a permutation
group, $\S_n$, at all, but by the cyclic group $\Integers_n$.  With this new
type of group comes a new type of statistics, in which a cyclic permutation
of the geons produces the complex phase $q$ or $\bar{q}$, $q=1^{1/n}$ being
an $n$th-root of unity.

Another interesting possibility, arising when the 
UIR, $\Gamma$, of $(slides)$ is non-abelian,  is that of ``projective
statistics,'' meaning a type of statistics expressed by a properly
projective representation of the permutation group or one of its
subgroups.  To construct such an 
example, we would need at least four geons because $\S_n$
possesses properly projective representations only for $n\ge{}4$.  

\subsection{No spin-statistics correlation}

All the UIRs of the MCG are on a equal footing as far as 
the canonical theory goes. Given a fixed 
spatial three-manifold, there appears in the theory 
a set of, in principle unpredictable, new parameters, some discrete
and some continuous, corresponding to Nature's choice of a 
particular UIR amongst all the possibilities.  
Some of these UIRs violate the usual spin-statistics
correlation in the grossest possible way. Subsection 
\ref{trivslides.subsection} shows that 
if the basic prime  is tensorial (the $2\pi$ rotation is 
trivial) then there is a UIR in which the permutation group 
is represented by the totally antisymmetric UIR. If the basic 
prime is spinorial then the quantum geon can be tensorial 
or spinorial and the permutation group be represented 
either by the totally symmetric or antisymmetric UIR.
In other words there are UIRs in which the geons are  
spinorial bosons 
or tensorial fermions.  

This lack of a correlation can be attributed to the fact that
in a theory in which the topology is fixed (such as canonical quantization)
 there is no
allowance for geon-anti-geon production. This is because 
a process in which a geon
and anti-geon are created from $\Reals^3$ is a topology changing one (we know this
from the decomposition theorem: one piece of non-trivial topology can't
``cancel'' another). The known spin-statistics theorems
for objects such as skyrmions and other kinks which have these emergent
properties of spin and statistics all require, for their proofs, that the
process of pair creation and annihilation be describable as a path in
the configuration space (\cite{Sorkin:1988ud} is a general such proof).
This leads one to expect that in a formulation of quantum gravity in
which topology change is naturally accommodated the spin
statistics correlation might be recovered. The SOH is such a formulation 
and  
we therefore turn now to that approach.

\section{Spacetime approach}

In \cite{Dowker:1996ei} it was shown that two non-chiral geons ({\it i.e.},
one which are their own antiparticle) which are pair created in a topology 
changing process, satisfy a limited spin-statistics theorem which eliminates
some of the spin-statistics violating sectors. Instead of reviewing
that work, we will turn to the more general question of how a 
general spin-statistics
theorem could be proved for all geons.   

In the SOH framework the
fundamental dynamical input is a rule attaching a quantum amplitude to each
pair of truncated histories which ``come together'' at some ``time''
\cite{Bialynicki-Birula,Sorkin:1994cd,Daughton:1994,Sorkin:1994}.  
Let us call such a pair a ``Schwinger history''
and its underlying manifold a ``Schwinger manifold''.  In the
case of quantum gravity, a truncated history is a Lorentzian manifold with
final boundary
(and possibly initial boundary depending on the physical
context), and the ``coming together'' means the identification or ``sewing
together'' of the final boundaries.  
In a framework in which topology change is allowed, all 
four-manifolds and all metrics on each manifold are summed over. 
Now, without disturbing the classical limit of the theory or the local
physics, we can multiply the amplitude of each Schwinger history by a
``weight'' $w$ depending only on the topology of the underlying
manifold (and on the two initial metrics, if initial boundaries are
present).  We would like, somewhat analogously to \cite{Laidlaw:1971ei},
to argue that
these weights carry some unitary
representation of $G$, and that sets of weights belonging to disjoint
representations ``do not mix''.  
We cannot do so in the present case for several reasons. We do not know
what $G$ is in a topology changing scenario precisely because the topology 
of space changes and so the MCG changes with it. Even if we could 
overcome this, we would be faced with the question of how to implement 
the higher-dimensional non-abelian UIR's. 

We will ignore these detailed questions here and instead sketch what a 
spin-statistics theorem might look like in a SOH framework.

\begin{figure}
\centerline{\epsfig{file=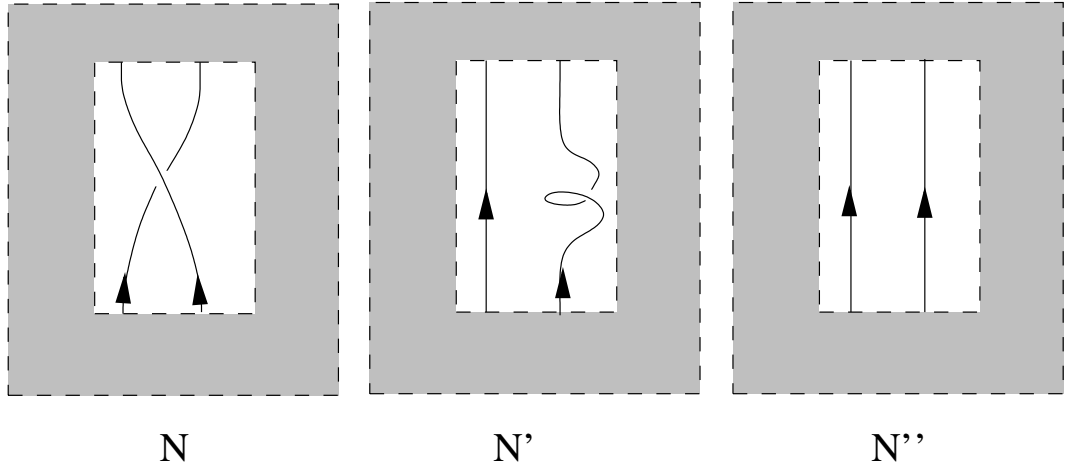}}
\caption{Three manifolds which differ only in the
unshaded compact region}
\label{en.fig}
\end{figure}

Consider two Schwinger manifolds, $N$ and $N'$, 
that contribute to the 
SOH and which differ only in a compact region in which,
in $N$ two  
identical primes swap places and in $N'$ one prime stays put
and the other rotates around by $2\pi$. This is illustrated in 
figure \ref{en.fig} in which the ``environment'' depicted in grey
is common to 
$N$ and $N'$. 
The lines represent the ``worldlines'' of the 
primes and are in reality framed curves in a four-dimensional 
manifold. (Given a framed curve in a four-dimensional space, there
is a topologically unique way of attaching a prime to each point of that 
curve.) The corkscrew effect in $N'$ represents the 
frame rotating by $2\pi$. 
 
The geons are bosons (fermions) if the weight of 
$N$ is equal to (minus) the weight of a third manifold, $N''$ which 
differs from $N$ only in that the two primes do not move at all
in that same compact region. 
The geons are tensorial (spinorial) if the weight of
$N'$ is equal to (minus) the weight of $N''$. 
A spin statistics theorem would consist of the result that 
given any two Schwinger manifolds such as $N$ and $N'$, 
they must have the same weight 
in the SOH. 

The rules that would achieve this result are the following 
\cite{Sorkin:1988mw}.
We dub manifolds with the same weight ``congruent.''

\noindent {$\bullet$}\ 
Manifolds which are diffeomorphic, via a diffeomorphism which
is the identity on the initial boundaries and at spatial infinity if these
are present, are congruent.

\noindent {$\bullet$}\
Let two manifolds, $N_1$ and $N_2$  differ only in a certain 
``compact region,'' so $N_1 = R_1 \cup E$ and $N_2 = R_2 \cup E$,
where $E$ is the ``environment''. If $N_1$ and $N_2$ are congruent then 
 $N_1' = R_1 \cup E'$ and $N_2' = R_2 \cup E'$ are also congruent.  

\noindent {$\bullet$}\ 
If $N$ is a manifold which admits a spacelike hypersurface
 which 
divides $N$ in half so that the second half is the time reverse of the 
first, then $N$ is congruent to the product manifold $V_0 \times [0,1]$
where $V_0$ is the initial boundary of $N$. (If there's no initial 
boundary then $N$ is congruent to $S^4$). 

\begin{figure}
\centerline{\epsfig{file=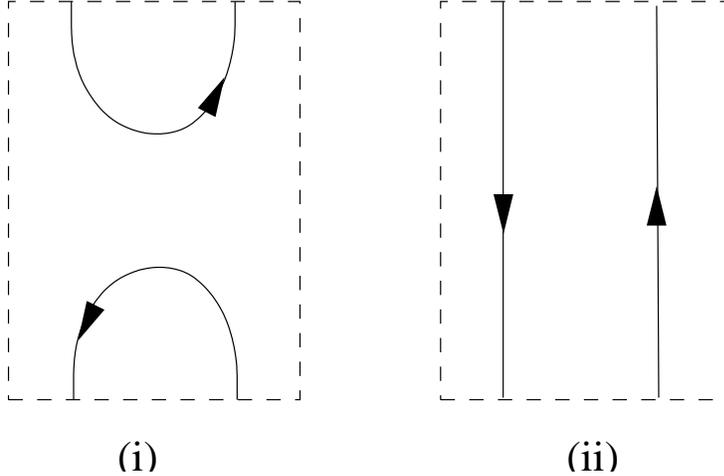}}
\caption{The annihilation of a prime-anti-prime pair 
followed by a pair creation
is congruent to the trivial propagation in time of a prime-anti-prime
pair}
\label{pair.fig}
\end{figure}

Now, for any geon topology, $P$, there is a canonical manifold, $X$, which 
mediates geon-anti-geon
annihilation. The third rule has the 
consequence that the manifold which consists of $X$ followed by 
its time reverse (pair annihilation followed by pair creation) 
is congruent to the manifold in which a prime and an anti-prime
just sit there. This is illustrated in figure
\ref{pair.fig} where the lines again represent the
worldlines of the primes and the arrows remind us that 
they are framed curves. The ``anti-prime'' 
is the mirror image, or ``chiral conjugate,'' 
of the prime.   

\begin{figure}
\centerline{\epsfig{file=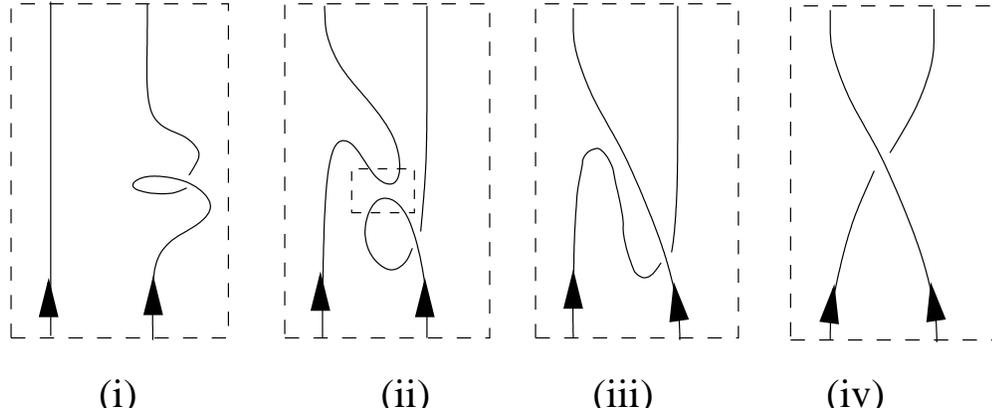}}
\caption{A sequence of congruent manifolds which establish a spin-statistics
correlation for geons}
\label{spinstat.fig}
\end{figure}

This can be used  
to prove a spin-statistics theorem 
as illustrated 
in figure \ref{spinstat.fig} which shows a sequence of 
congruent compact regions, to be thought of as embedded
in some common environment. Again, the lines represent the 
(framed) worldlines of two identical primes. Region (i) 
contains a static prime and one which rotates by $2\pi$. ``Untwisting''
the rotation gives region (ii) which is diffeomorphic to (i).
Notice that region (ii) contains 
a small region which is a pair-annihilation-pair-creation
event and replacing that with the prime-anti-prime product, 
as we may do by rule 3, 
gives region (iii). This is in turn diffeomorphic to (iv). 
So (i) is congruent to (iv) which is the statement of 
a spin-statistics correlation in this framework. 

\begin{figure}
\centerline{\epsfig{file=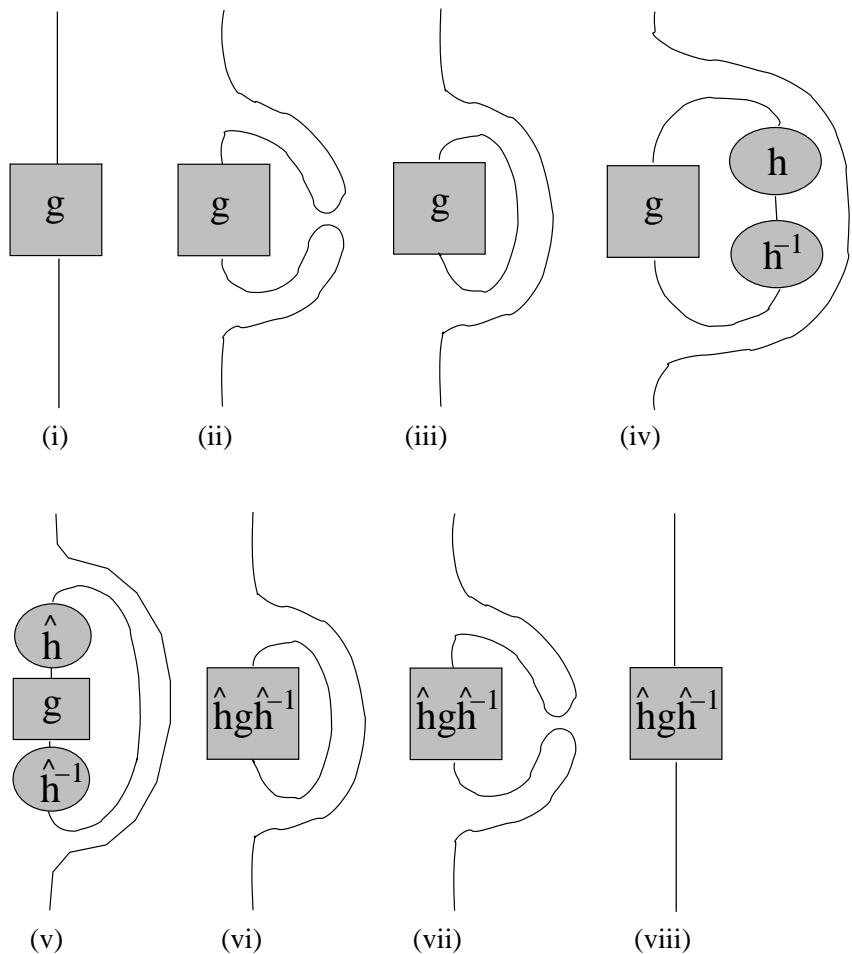}}
\caption{abelian proof}
\label{smallabelian.fig}
\end{figure}

It has now been realised that these rules are much more powerful 
than originally envisaged. In particular they allow only  
abelian UIR's as potential weights and thus eliminate the 
possibility of parastatistics. (This seems to be a  
consequence of ``topological'' spin-statistics theorems in 
general \cite{Balachandran:1990wr,Balachandran:1993mf}.)
A sketch of the 
proof of this result is given in figure \ref{smallabelian.fig}
which shows a sequence of (compact regions of)
congruent manifolds  
-- the surrounding ``boxes'' are to be imagined.

In (i) is shown 
a manifold in which at early times there are a number
of primes sitting in fixed positions, propagating 
in time: that's denoted by the single 
vertical line which now stands for a collection of 
a number of framed worldlines. 
In the intermediate region a large diffeo, $g$, 
an element of the MCG of the three-manifold comprising those
prime summands, is developed: that's denoted by the shaded 
box labeled $g$. It's useful to imagine 
an example in which, say, there are two identical primes and 
$g$ is the exchange diffeo so that  diagram (i) is actually
the manifold $N$ in figure \ref{en.fig}. (i) is diffeomorphic
to (ii) which is congruent to (iii) by rule 3. The line on the 
right in (iii) is now a spectator for the next few steps. Into 
the ``closed loop'' we may introduce the development of any
diffeo $h$ followed by its inverse. 
$h$ is an element of the MCG not of the original multi-prime 
manifold but one which is its mirror image in which each 
original prime is replaced by its chiral conjugate.
This can be seen by following carefully what is created and destroyed
at the pair creation and annihilation events.  
Then the diffeos $h$ and $h^{-1}$ are eased 
along the loop in opposite directions until they are adjacent 
to $g$ as shown in (v). In doing so $h$ becomes  a new diffeo, $\hat h$,
an element in the same MCG as $g$, related to $h$ in the obvious way.
Similarly for ${\hat h}^{-1}$. Composing ${\hat h}$, 
$g$ and ${\hat h}^{-1}$ gives (vi) and the reverse of the first two 
steps results in (viii). This may be done for any $h$ and so 
(i) is congruent to (viii) for any $\hat h$ in the MCG. 

This means that if we are able, in some effective sense,
to implement the Laidlaw-Morette-DeWitt type of scheme, the weight of 
manifold (i) would have to be the same as that of manifold
(viii). Thus, the weights can only depend on the conjugacy class of
the elements of the mapping class group and so can only 
carry abelian UIR's.  

It might be seen as an advantage that only abelian UIR's need be
considered -- the representation theory is radically simplified in this
case and we feel that we understand how to attach mere phases as
weights to the manifolds in the SOH. But there is a potential, serious
drawback. All spinorial geons known to us are non-abelian. 
That is, we do not know of a prime whose $2\pi$ rotation is 
non-trivial in an abelian UIR of the MCG. If no such primes do in 
fact exist, the restriction to abelian UIR's would eliminate 
spinorial and fermionic geons altogether and this seems much too 
high a price to pay for a spin-statistics theorem.

\section{Motivation}

Why should we care whether or not topological geons obey a 
spin-statistics theorem? We can hope that by formulating  
rules that would give us a spin-statistics theorem we may glean 
clues about the nature of the underlying, more fundamental theory
of quantum gravity 
which should eventually give rise to such rules. It is clear that the  
spin-statistics question is intimately connected to the 
question of the weights attachable to the different manifolds in
the SOH, whose arbitariness in the absence of something like these
ideas 
seems a problem. 
Geons contain the potential to be the mechanism of the ultimate 
unification between spacetime and matter: perhaps spacetime is all there
is. Bosonic fields could arise from Kaluza-Klein reductions 
and fermions could be quantum geons. Indeed Friedman and Higuchi showed 
that in pure Kaluza-Klein gravity there will be stable geons with
the kinematical quantum numbers of the standard model, 
although obtaining chiral fermions is problematic \cite{Friedman:1990az}.
(Showing that
geons have the standard model 
masses is a problem of a different order,
though one potential difficulty has 
been slightly alieviated since the discovery of a neutrino mass. There is 
still an enormous number, $m_{planck}/m_{neutrino}$, to explain but at least 
it's finite.) As unlikely as this might seem, the   
payoff, were it to be true, is so high that it is worth bearing the
possibility in mind.  Then a geon spin-statistics
theorem would give us the fundamental explanation 
of the perfect spin-statistics correlation we see in nature. 

\section{Acknowledgments} 
R.D.S would like to thank the Aspen Center for Physics for its
hospitality and to acknowledge support from NSF grant PHY-9600620.  
H.F.D. is supported in part by an EPSRC Advanced Fellowship.


\begin{references}


\bibitem{Friedman:1980st}
Friedman, J.L.,  and Sorkin, R.D.,
\newblock {\em Phys. Rev. Lett.} {\bf44}, 1100 (1980).

\bibitem{Friedman:1982du}
Friedman, J.L.,  and Sorkin, R.D.,    
\newblock {\em Gen. Rel. Grav.} {\bf 14}, 615 (1982).

\bibitem{Sorkin:1985bg}
Sorkin, R.D., 
\newblock ``Introduction to topological geons,''
\newblock in {\em Proceedings of the
  NATO Advanced Study Institute on Topological Properties and Global Structure
  of Space-Time, Erice, Italy, May 12-22, 1985},
edited by  P.G. Bergmann and V.~De Sabbata, Plenum Press, New York, 1986. 

\bibitem{Sorkin:1988mw}
Sorkin, R.D.,  
\newblock ``Classical topology and quantum phases: Quantum geons,''
\newblock in {\em
  Proceedings, Geometrical and algebraic aspects of nonlinear field theory,
  Amalfi, Italy, May 1988}, edited by 
G.~Marmo, S.~De~Filippo, M.~Marinaro and G.~Vilasi, North Holland, Amsterdam, 
New York, 1989. 

\bibitem{Aneziris:1989xz}
Aneziris, C., et~al,
\newblock {\em Mod. Phys. Lett.} {\bf A4}, 331 (1989).

\bibitem{Aneziris:1989cr}
Aneziris, C., et~al,
\newblock {\em Int. J. Mod. Phys.} {\bf A4}, 5459 (1989).

\bibitem{Sorkin:1996yt}
Sorkin, R.D.,  and Surya, S.,
\newblock {\em Int. J. Mod. Phys.} {\bf A13}, 3749 (1998).

\bibitem{Sorkin:1996cv}
Sorkin, R.D.,  and Surya, S., 
\newblock ``Geon statistics and UIR's of the mapping class group,''
\newblock in 
{\em Proceedings: First Latin American 
Symposium on High Energy Physics and
   VII Mexican School of Particles and Fields, M{\'e}rida, M{\'e}xico,
Nov. 1996}, edited by 
Juan Carlos D'Olivo, Martin Klein-Kreislev, H{\'e}ctor M{\'e}ndez,
AIP Conference Proceedings 400, New York,   1997.

\bibitem{Balachandran:1990wr}
Balachandran, A.P., et~al,
\newblock {\em Mod. Phys. Lett.} {\bf A5}, 1575--1586 (1990).

\bibitem{Balachandran:1993mf}
Balachandran, A.P., et~al, 
\newblock {\em Int. J. Mod. Phys.} {\bf A8}, 2993--3044 (1993).

\bibitem{Finkelstein:1968hy}
Finkelstein, D.,  and Rubinstein, J.,
\newblock {\em J. Math. Phys.} {\bf 9}, 1762 (1968).

\bibitem{Dowker:1996ei}
Dowker, H.F.,  and Sorkin, R.D.,
\newblock {\em Class. Quant. Grav.} {\bf 15},  1153 (1998).

\bibitem{Sorkin:1994cd}
Sorkin, R.D.,
\newblock {\em Int. J. Theor. Phys.} {\bf 33}, 523--534 (1994).


\bibitem{Friedman:1988qs}
Friedman, J.L.,
\newblock ``Space-time topology and quantum gravity,''
\newblock in {\em Conceptual problems of
  quantum gravity: Proceedings of the 1988 Osgood Hill Conference, North
  Andover, Massachusetts, 15-19 May 1988}, 
edited by A.~Ashtekar and J.~Stachel, Birkh\"auser, Boston, 1991. 

\bibitem{Giulini:1994de}
Giulini, D.,
\newblock {\em Int. J. Theor. Phys.} {\bf 33}, 913--930 (1994).

\bibitem{Giulini:1995ui}
Giulini, D.,
\newblock {\em Helv. Phys. Acta} {\bf 68}, 86--111 (1995).

\bibitem{Giulini:1997hi}
Giulini, D.,
\newblock {\em Banach Center Publ.} {\bf 39}, 303--315 (1997). 

\bibitem{Sorkin:1988ud}
Sorkin, R.D., 
\newblock {\em Commun. Math. Phys.} {\bf 115}, 421 (1988).

\bibitem{Bialynicki-Birula}
Bialynicki-Birula, I.,
\newblock ``Transition amplitudes versus transition probabilities and a
     reduplication of space-time''
\newblock in {\em 
  Quantum Concepts in Space and Time}, edited by R. Penrose and C.J. Isham,
Clarendon Press,  Oxford, 
  1986.  

\bibitem{Daughton:1994}
Louko, J.,  Daughton, A., and Sorkin, R.D.,
\newblock In {\em Proceedings of the
  5th Canadian Conference on General Relativity and Relativistic Astrophysics},
  edited by R.B. Mann and R.G. McLenaghan, World Scientific, 
  Singapore, 1994.

\bibitem{Sorkin:1994}
Sorkin, R.D., 
\newblock ``Quantum measure theory and its interpretation,''
\newblock in {\em Quantum Classical
  Correspondence: Proceedings of the 4th. Drexel Symposium on Quantum
  Nonintegrability, Philadelphia, USA, September 8-11, 1994}, 
edited by D.H. Feng and B-L Hu, International Press,  Cambridge Mass.,
  1997. 

\bibitem{Laidlaw:1971ei}
Laidlaw, M.G.G., and Morette-DeWitt, C.,
\newblock {\em Phys. Rev.} {\bf D3}, 1375--1378 (1971).

\bibitem{Friedman:1990az}
Friedman, J.L.,  and Higuchi, A.,
\newblock {\em Nucl. Phys.} {\bf B339}, 491 (1990).

\end{references}
\end{document}